\documentclass[prb,twocolumn,showpacs]{revtex4}% Physical Review B

\usepackage{graphicx}% Include figure files
\usepackage{dcolumn}% Align table columns on decimal point
\usepackage{bm}% bold math

\emergencystretch=\maxdimen
\begin{document}

%\preprint{APS/123-QED}

\title{Spectral properties near the Mott transition in the two-dimensional $t$-$J$ model}

\author{Masanori Kohno}
\affiliation{WPI Center for Materials Nanoarchitectonics, 
National Institute for Materials Science, Tsukuba 305-0044, Japan}

\date{\today}

\begin{abstract}
The single-particle spectral properties of the two-dimensional $t$-$J$ model in the parameter regime relevant to cuprate high-temperature superconductors are investigated using cluster perturbation theory. 
Various anomalous features observed in cuprate high-temperature superconductors are collectively explained 
in terms of the dominant modes near the Mott transition in this model. 
Although the behavior of the dominant modes in the low-energy regime is similar to that in the two-dimensional Hubbard model, 
significant differences appear near the Mott transition for the high-energy electron removal excitations which can be considered to primarily originate from holon modes in one dimension. 
The overall spectral features are confirmed to remain almost unchanged as the cluster size is increased from $4\times 4$ to $6\times 6$ sites 
by using a combined method of the non-Abelian dynamical density-matrix renormalization group method and cluster perturbation theory. 
\end{abstract}

\pacs{71.30.+h, 71.10.Fd, 74.72.Kf, 79.60.-i}

\maketitle
\section{Introduction} % ------ Introduction ------
Cuprate high-temperature (high-$T_{\rm c}$) superconductors, which are obtained by doping Mott insulators containing CuO$_2$ planes, \cite{highTc} 
exhibit various features that appear anomalous from conventional viewpoints. \cite{ShenRMP,Graf,kink_Xie,kink_Valla,kink_Pan,kink_Moritz,UniversalFlatbandBi2212,LSCO_FS,XraySWT_PRB,XraySWT_PRL,DagottoRMP,ImadaRMP} 
Because the anomalous features are considered to be related to high-$T_{\rm c}$ superconductivity, 
the effects of electronic correlations near the Mott transition in two-dimensional (2D) systems have attracted considerable attention. \cite{AndersonRVB,DagottoRMP,ImadaRMP,NagaosaRMP,OgataRev,YanaseRev} 
In particular, through an analysis of electronic correlations among relevant Cu and O orbitals, the 2D $t$-$J$ model has been derived as a minimal model of high-$T_{\rm c}$ cuprates. \cite{ZhangRice} 
However, its properties and their relationship to the anomalous features are not well understood, primarily because of difficulties in dealing with the no-double-occupancy constraint. 
\par
Although this model can also be derived effectively from the 2D Hubbard model in the large-repulsion limit, \cite{HarrisUexp} 
it is not clear how similar the two models' properties are in the parameter regime relevant to high-$T_{\rm c}$ cuprates. 
In fact, the 2D $t$-$J$ and Hubbard models have frequently been studied from different viewpoints: 
the former has been considered a doped Mott insulator accessible from slave-particle theories, \cite{NagaosaRMP,OgataRev} 
while the latter has been considered a strongly interacting electron system accessible from Fermi liquid theory. \cite{YanaseRev} 
In some studies, double occupancy, which is excluded in the $t$-$J$ model, has been considered important 
to the anomalous features. \cite{PhillipsRMP,PhillipsMottness,SakaiImadaPRL,ImadaCofermionPRL,ImadaCofermionPRB} 
\par
In this paper, by using cluster perturbation theory (CPT), \cite{CPTPRL,CPTPRB} similarities and dissimilarities in the spectral features of these models are clarified. 
In addition, various anomalous features observed in high-$T_{\rm c}$ cuprates are collectively explained in the 2D $t$-$J$ model, 
which is an effective model of the CuO$_2$ plane and has no double occupancy. 
A method to reduce cluster-size effects is also introduced. 
\section{Models and Methods} % ------ Models and Methods ------
\subsection{Models} % ------ Models ------
We consider the 2D $t$-$J$ model defined by the following Hamiltonian for $t>0$ and $J>0$: 
$$%\begin{equation}
%{\cal H}=-t\sum_{\langle i,j\rangle\sigma}({\tilde c}_{i,\sigma}^{\dagger}{\tilde c}_{j,\sigma}+{\mbox {H.c.}})+J\sum_{\langle i,j\rangle}\left({\bm S}_i\cdot{\bm S}_j-\frac{1}{4}n_in_j\right)-\mu\sum_in_i, 
{\cal H}=-t\sum_{\langle i,j\rangle,\sigma}({\tilde c}_{i,\sigma}^{\dagger}{\tilde c}_{j,\sigma}+{\mbox {H.c.}})+J\sum_{\langle i,j\rangle}({\bm S}_i\cdot{\bm S}_j-\frac{1}{4}n_in_j)-\mu\sum_in_i, 
$$%\end{equation}
where ${\tilde c}_{i,\sigma}=(1-n_{i,-\sigma})c_{i,\sigma}$ and $n_i=\sum_\sigma n_{i,\sigma}$ for the annihilation operator $c_{i,\sigma}$ and number operator $n_{i,\sigma}$ of an electron with spin $\sigma$ at site $i$. 
Here, ${\bm S}_i$ denotes the spin operator at site $i$, and $\langle i, j\rangle$ means that sites $i$ and $j$ are nearest neighbors on a square lattice. 
At half-filling (doping concentration $\delta=0$), the model reduces to the Heisenberg model. 
The $t$-$J$ model can also be effectively obtained for $J=4t^2/U$ by neglecting the three-site term \cite{HarrisUexp,OgataRev} in the large-$U/t$ limit of the Hubbard model defined by the following Hamiltonian: 
$$
{\cal H}_{\rm Hub}=-t\sum_{\langle i,j\rangle,\sigma}(c_{i,\sigma}^{\dagger}c_{j,\sigma}+{\mbox {H.c.}})+U\sum_in_{i,\uparrow}n_{i,\downarrow}-\mu\sum_in_i.
$$ 
\par
\begin{figure*}
\includegraphics[width=17.8cm]{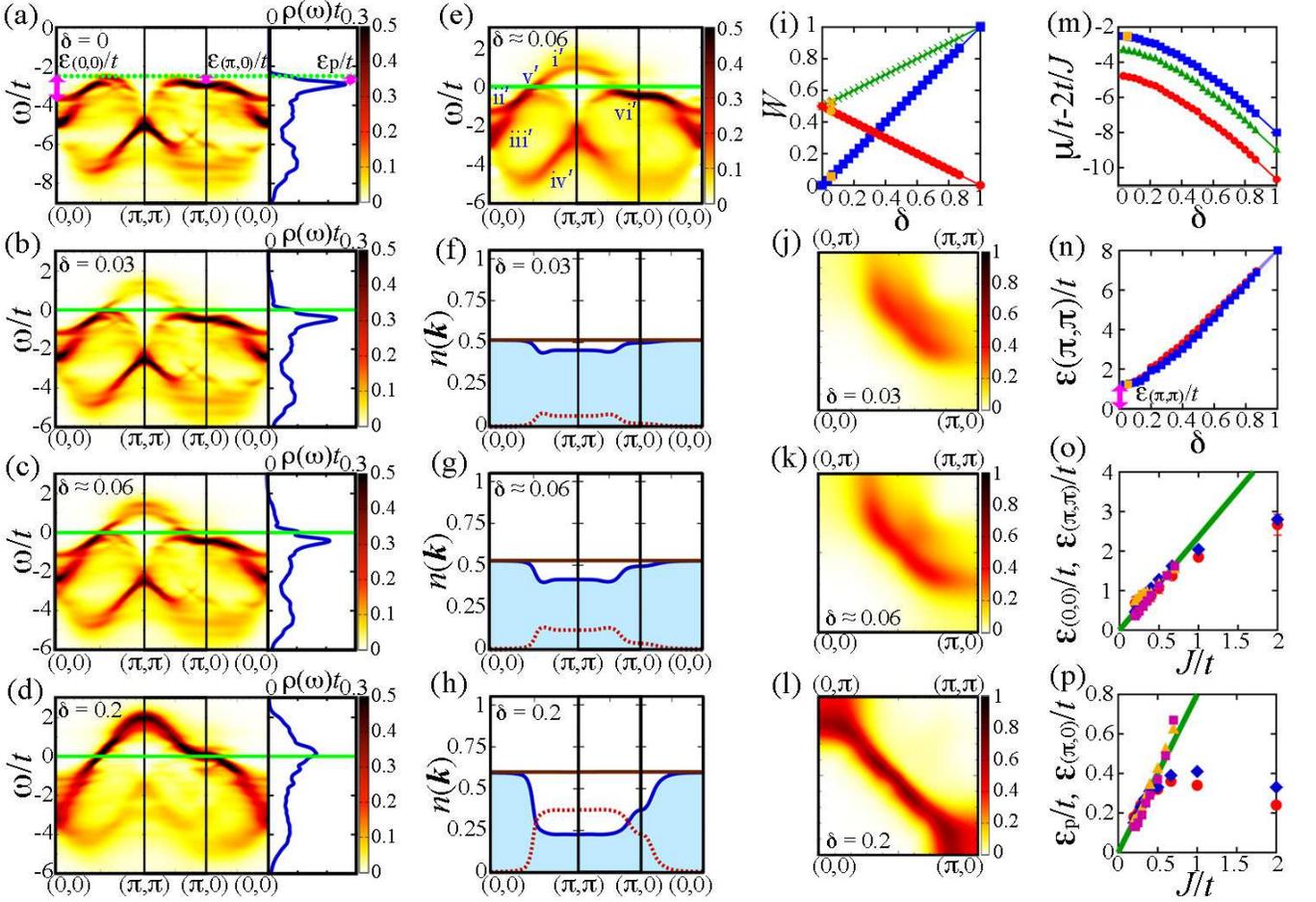}
\caption{Results for the 2D $t$-$J$ model obtained using $(4\times 4)$-site CPT [(a)--(d) and (f)--(p)] and $(6\times 6)$-site CPT [(e); yellow symbols in (i), (m), and (n) for $J/t=0.5$ at $\delta=1/18$]. 
(a)--(d) $A({\bm k}, \omega)t$ for $J/t = 0.5$ at $\delta = 0$ ($\mu/t=4$), 0.03, $1/18(\approx 0.06)$, and 0.2 (from above). The rightmost panels show $\rho(\omega)t [\equiv \int d{\bm k}A({\bm k}, \omega)t/(2\pi)^d]$ in $d(=2)$ dimensions. 
The dotted green line in (a) indicates $\omega$ at the top of the band. The solid green lines in (b)--(d) indicate $\omega = 0$. 
(e) The same as the left panels of (c) but using $(6\times 6)$-site CPT. The blue numbers indicate dominant modes. 
(f)--(h) $n({\bm k}) [\equiv\int_{\omega\,{\rm range}} d\omega A({\bm k}, \omega)]$ for (b)--(d) for $\omega < 0$ [$n_-({\bm k})$] (solid blue curves with hatches), $\omega > 0$ (dotted red curves), and all $\omega$ $[n_{\rm t}({\bm k})$] (solid brown lines). 
(i) Spectral weight $W[\equiv \int_{\omega\,{\rm range}} d\omega\rho(\omega)]$ for $\omega < 0$ [$W_-$] (red circles), $\omega > 0$ [$W_+$] (blue squares), and all $\omega$ (green crosses) at $J/t = 0.5$. 
(j)--(l) The same as in (b)--(d) but for $\omega\approx 0$. 
(m) Chemical potential $\mu$ for $J/t=0.5$, 0.4, and 0.3 (from above). 
(n) Energy of mode i$^{\prime}$ at $(\pi,\pi)$ [$\varepsilon(\pi,\pi)$] (blue squares) for $J/t = 0.5$. 
%The yellow symbols in (f), (g), and (n) indicate the results obtained using $(6\times 6)$-site CPT. 
(o) $\varepsilon_{(0,0)}$ (purple squares) and $\varepsilon_{(\pi,\pi)}$ (yellow triangles). 
The solid green line indicates $\sqrt{2}v_{\rm s}/t$ ($v_{\rm s}\approx 1.18\sqrt 2 J$ [\onlinecite{Singh}]). 
(p) $\varepsilon_{\rm p}$ (purple squares) and $\varepsilon_{(\pi,0)}$ (yellow triangles). 
The solid green line is proportional to $J$. 
The red circles in (n) and blue diamonds and red circles in (o) and (p) indicate corresponding energies in the 2D Hubbard model ($U/t=4t/J$), taken from Ref. \onlinecite{Kohno2DHub}. 
Gaussian broadening with a standard deviation of $0.1t$ is used.}
\label{fig:2DtJ}
\end{figure*}

In the ground state near the Mott transition, ferromagnetic fluctuations might be dominant in the very small-$J/t$ regime \cite{Nagaoka,PutikkaFM,ZhangFM} and 
phase separation occurs in the large-$J/t$ regime. \cite{DagottoRMP,EmeryPS1,PutikkaPS1,Kohno2DtJGS,HellbergPS,ShihPS1,DagottoPS2,Dagotto_sumrule,SorellaPS1, WhitePS,OgataRev} 
Here, focusing attention on the parameter regime relevant to high-$T_{\rm c}$ cuprates ($J/t\approx 0.5$), \cite{DagottoRMP,OgataRev} 
where the CPT results exhibit no indication of phase separation [Fig. \ref{fig:2DtJ}(m)], 
we study the spectral function defined as $A({\bm k}, \omega)=-\mbox{Im}G({\bm k}, \omega)/\pi$. 
Here, $G({\bm k}, \omega)$ denotes the retarded single-particle Green function for momentum ${\bm k}$ and frequency $\omega$ at zero temperature. \cite{ImadaRMP,DagottoRMP} 
%\begin{widetext}
%\end{widetext}
\subsection{Methods} % ------ Methods ------
In this paper, CPT is employed: $G({\bm k}, \omega)$ is calculated by connecting cluster Green functions through the first-order hopping process. \cite{CPTPRL,CPTPRB,CPTtJ,Kohno2DHub,Kohno2DHubNN,CPTJ3} 
By considering superclusters, $G({\bm k}, \omega)$ at arbitrary $\delta$ can be obtained. \cite{CPTPRB} 
In the large-cluster limit, CPT becomes exact. 
Here, to calculate the ($4\times4$)-site and ($6\times6$)-site cluster Green functions, exact diagonalization and the dynamical density-matrix renormalization group (DDMRG) method \cite{DDMRG} are used, respectively. 
In the DDMRG calculation, the U(1)$\otimes$SU(2) basis \cite{nonAbeliantJ,nonAbelianThesis} is employed \cite{KohnoDIS} and 1000 density-matrix eigenstates are retained. 
\par
The DDMRG method and CPT work well together because of the following reasons. (1) It is not necessary to repeat DDMRG calculations because CPT does not impose self-consistency. 
(2) Data under open boundary conditions, which are obtained accurately in the DDMRG method, are used in CPT. 
The combined method can be regarded either as CPT where the DDMRG method is utilized as a cluster Green function solver or as the DDMRG method where momenta are interpolated based on CPT. 
Note that real-space cluster Green functions are used in the combined method [Fig. \ref{fig:2DtJ}(e)]
in contrast with the random-phase approximation (RPA) from the decoupled-chain limit \cite{RPAWen} using DDMRG results [Figs. \ref{fig:1DtJ}(c) and \ref{fig:1DtJ}(f)]. \cite{Kohno2DHub} 
In this paper, the RPA from the decoupled-chain limit, which corresponds to the perturbation theory up to the first order with respect to interchain hopping, 
is only used to explain how the spectral weights are shifted by interchain hopping from the decoupled-chain limit and 
to trace the origins of the dominant modes of 2D systems back to those of one-dimensional (1D) systems (Secs. \ref{sec:domonantModes} and \ref{sec:largeNegative}). 
\subsection{Cluster-size effects} % ------ Cluster-size effects ------
As shown in Figs. \ref{fig:2DtJ}(c), \ref{fig:2DtJ}(e), \ref{fig:2DtJ}(i), \ref{fig:2DtJ}(m), and \ref{fig:2DtJ}(n), the results obtained using $(6\times 6)$-site CPT are almost the same as those 
using $(4\times 4)$-site CPT. 
This implies that the overall spectral features will not change significantly if the cluster size is increased. 
\section{Spectral properties} % ------ Spectral properties ------
\subsection{Dominant modes} % ------ Dominant modes ------
\begin{figure}
\includegraphics[width=8.4cm]{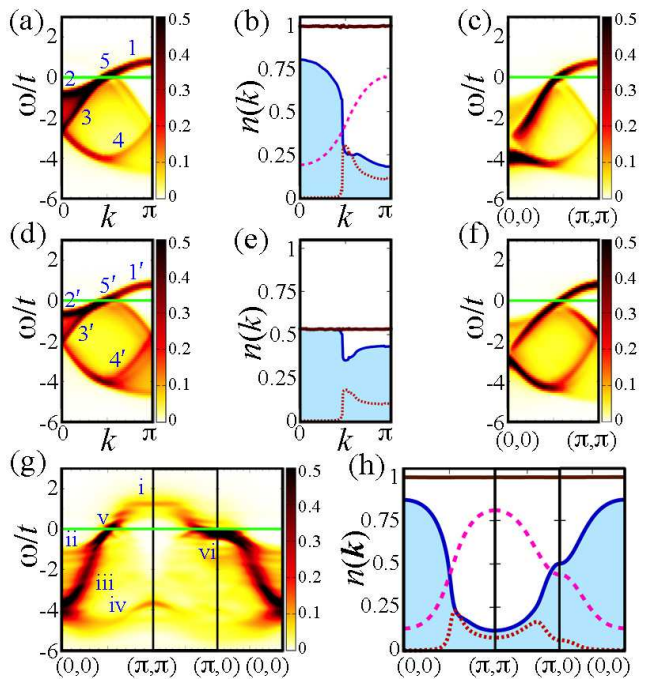}
\caption{(a) $A(k, \omega)t$ for $U/t = 8$ at $\delta=1/15(\approx 0.07)$ for the lower Hubbard band (LHB) of the 1D Hubbard model \cite{Kohno1DHub} on a 120-site chain 
obtained using the non-Abelian DDMRG method with 120 density-matrix eigenstates. 
The blue numbers indicate dominant modes. 
(b) $n(k)$ for (a) for $\omega < 0$ [$n_-(k)$] (solid blue curve with hatches), $\omega > 0$ in the LHB (dotted red curve), the upper Hubbard band (dashed pink curve), 
and all $\omega$ [$n_{\rm t}(k)$] (solid brown line). \cite{Kohno1DHub} 
(c) RPA results for the interchain hopping integral $t_{\perp} = t$ obtained using the data in (a). \cite{Kohno2DHub} 
(d)--(f) The same as (a)--(c) but for the $t$-$J$ model at $J/t=0.5$. 
(g) The same as the left panels of Fig. \ref{fig:2DtJ}(c) but for the LHB of the 2D Hubbard model at $U/t = 8$. \cite{Kohno2DHub} 
The blue numbers indicate dominant modes. 
(h) $n({\bm k})$ for (g) with the same line types as in (b). \cite{Kohno2DHub} 
The solid green lines indicate $\omega = 0$. Gaussian broadening with a standard deviation of $0.1t$ is used.}
\label{fig:1DtJ}
\end{figure}

\label{sec:domonantModes}
The results for the 2D $t$-$J$ model obtained using CPT are shown in Fig. \ref{fig:2DtJ}. 
The overall spectral features can be explained in terms of the dominant modes [Fig. \ref{fig:2DtJ}(e), modes i$^{\prime}$--vi$^{\prime}$] corresponding to those of the 2D Hubbard model [Fig. \ref{fig:1DtJ}(g), modes i--vi], \cite{Kohno2DHub} 
whose origins can be traced back to those of the 1D models [Fig. \ref{fig:1DtJ}(d), modes 1$^{\prime}$--5$^{\prime}$; Fig. \ref{fig:1DtJ}(a), modes 1--5] \cite{Kohno1DHub,SchulzSpctra,EsslerBook,DDMRGAkw,Shadowband,1dtJUinf,antiholon,BaresBlatter,EderOhtaShadowband} 
by considering how the spectral weights are shifted by interchain hopping from the decoupled-chain limit [Figs. \ref{fig:1DtJ}(c) and \ref{fig:1DtJ}(f); Sec. \ref{sec:largeNegative}]. \cite{Kohno2DHub,Kohno2DHubNN,KohnoNP,KohnoQ1DH} 
The dispersing mode around $(\pi,\pi)$ for $\omega>0$, mode i$^{\prime}$ ($\approx$ mode i), originates from mode 1$^{\prime}$ ($\approx$ mode 1, upper edge of the spinon-antiholon continuum). 
The mode around $(0,0)$ for $\omega\lesssim 0$, mode ii$^{\prime}$ ($\approx$ mode ii), primarily originates from mode 2$^{\prime}$ ($\approx$ mode 2, spinon mode), 
and the mode for slightly lower $\omega$, mode iii$^{\prime}$ ($\approx$ mode iii), primarily originates from mode 3$^{\prime}$ ($\approx$ mode 3, holon mode). 
The mode spreading over the Brillouin zone in the large negative $\omega$ regime, mode iv$^{\prime}$ ($\approx$ mode iv), primarily originates from mode 4$^{\prime}$ ($\approx$ mode 4, holon mode called the shadow band). 
The mode bending back around $(\pi/2,\pi/2)$ for $\omega\gtrsim 0$ [$\approx$ upper edge of the band around $(\pi/2,\pi/2)$ at half-filling], mode v$^{\prime}$ ($\approx$ mode v), originates from mode 5$^{\prime}$ ($\approx$ mode 5, antiholon mode). 
The flat mode, mode vi$^{\prime}$ ($\approx$ mode vi), is dominant around $(\pi,0)$. 
\subsection{Positive $\omega$} % ------ w>0 ------
Mode i$^{\prime}$ corresponds to the doping-induced states 
observed in high-$T_{\rm c}$ cuprates and in theoretical calculations \cite{XraySWT_PRB,XraySWT_PRL,ImadaRMP,DagottoRMP,PhillipsRMP,PhillipsMottness,SakaiImadaPRL,SakaiImadaPRB,ImadaCofermionPRL,ImadaCofermionPRB,Kohno2DHub,Kohno2DHubNN,KohnoSpin,EderOhtaIPES,EderOhta2DHub,EderOhta_band,Ohta_ingapstates,Stephan_nk,Tohyama2DtJ,ZemljicFiniteT,PreussPG,PreussQP,Bulut,MoreoQP,JaklicAkw,JaklicAkwRev,Dagotto_sumrule,Eskes,DagottoDOS} 
with controversial interpretations. 
The CPT results indicate that the energy of this mode at $(\pi,\pi)$ [$\varepsilon(\pi,\pi)$] does not reach zero even in the small-doping limit [Fig. \ref{fig:2DtJ}(n)]. 
The extrapolated value of $\varepsilon(\pi,\pi)$ to $\delta\rightarrow 0$ [$\varepsilon_{(\pi,\pi)}$] behaves essentially as $\sqrt{2}v_{\rm s}$ in the small-$J/t$ regime, 
where $v_{\rm s}$ denotes the spin-wave velocity of the 2D Heisenberg model ($v_{\rm s}\approx 1.18\sqrt 2 J$ [\onlinecite{Singh}]) [Fig. \ref{fig:2DtJ}(o)], as in the 2D Hubbard model. \cite{Kohno2DHub} 
In addition, the spectral weight for $\omega>0$ ($W_+$) behaves exactly as $\delta$ [Fig. \ref{fig:2DtJ}(i)]. \cite{Stephan_nk,Dagotto_sumrule} 
These results imply that mode i$^{\prime}$ continuously leads to the magnetic excitation of the Mott insulator, while its spectral weight gradually disappears toward the Mott transition. 
%The above results suggest that mode i$^{\prime}$ continuously leads to the magnetic excitation of the Mott insulator, while its spectral weight gradually disappears toward the Mott transition. 
This feature is essentially the same as that in the 1D and 2D Hubbard models \cite{Kohno1DHub, Kohno2DHub,KohnoSpin} and is consistent with a general argument in the small-doping limit. \cite{KohnoDIS} 
Thus, this feature will be related to transformation to the Mott insulator, which has low-energy spin excitation but no low-energy charge excitation (spin-charge separation), rather than being related to double occupancy. 
\par
Because mode v$^{\prime}$ bends back around $(\pi/2,\pi/2)$ for $\omega\gtrsim 0$ [Figs. \ref{fig:2DtJ}(b), \ref{fig:2DtJ}(c), and \ref{fig:2DtJ}(e)], there are small intensities at momenta slightly away from $(\pi/2,\pi/2)$ toward $(\pi,\pi)$ as well as significant intensities around $(\pi/2,\pi/2)$ for $\omega\approx 0$ near the Mott transition [Figs. \ref{fig:2DtJ}(j) and \ref{fig:2DtJ}(k)]. \cite{Kohno2DHub} 
The region surrounded by these intensities might be effectively regarded as a hole pocket. 
Regarding hole pockets, the momentum distribution function $n_-({\bm k})$ has been investigated; \cite{DagottoRMP,Stephan_nk,Singh_nk,Putikka_nk,Ding_nk,EderOhta_nk,EderOhta_band,Haas_nk,Chernyshev_nk,Leung_nk} 
a high momentum resolution is desired near the Mott transition at zero temperature. 
The CPT results indicate that $n_-({\bm k})$ exhibits a small dip near $(\pi/2,\pi/2)$ [Figs. \ref{fig:2DtJ}(f) and \ref{fig:2DtJ}(g)], 
for which mode v$^{\prime}$ would be more or less responsible. \cite{KohnoDIS,EderOhtaIPES,EderOhta_band} 
%For precise determination of hole pockets, more accurate investigations around $\omega=0$ are necessary. 
\subsection{Small negative $\omega$} % ------ w<0 ------
The bandwidth of mode ii$^{\prime}$ at half-filling [$\varepsilon_{(0,0)}$] [Fig. \ref{fig:2DtJ}(a)] has been studied in relation to the hole motion in an antiferromagnet, 
and its $J$ dependence (particularly the power-law behavior in the small-$J/t$ regime) has been discussed. \cite{DagottoRMP,MoreoQP,JaklicAkwRev,Poilblanc1h,DagottotJ1h,DagottotJHub1h,KaneLeeRead1h,Sachdev1h,SchmittRink1h,TrugmanString,Hamer1h,Marsiglio1h,MartinezHorsch1h,Szczepanski1h,LiuManousakis1h,Brunner1h2D,Beran1h2D,CPTJ3,Nagaev1h} 
Figure \ref{fig:2DtJ}(o) indicates that it behaves essentially as $\sqrt{2}v_{\rm s}$ in the small-$J/t$ regime, 
as in the 2D Hubbard model. \cite{Kohno2DHub,Kohno2DHubNN,KohnoSpin} 
The bifurcation between this mode and mode iii$^{\prime}$ as well as the reduction in spectral weight just below the bifurcation point [Figs. \ref{fig:2DtJ}(b), \ref{fig:2DtJ}(c), and \ref{fig:2DtJ}(e)] 
can be identified \cite{Kohno2DHub} as the giant kink and waterfall behavior observed in high-$T_{\rm c}$ cuprates \cite{Graf,kink_Xie,kink_Valla,kink_Pan,kink_Moritz} 
for which various interpretations have been proposed. \cite{Graf,kink_Xie,kink_Valla,kink_Pan,kink_Moritz,Kohno2DHub,SakaiImadaPRB,PhillipsMottness,PhillipsRMP,ZemljicFiniteT,ScalapinoKink,kink_Tan,kink_Weber,AvellaCOM} 
Modes ii$^{\prime}$ and iii$^{\prime}$ correspond to spinon-like and holon-like branches observed in high-$T_{\rm c}$ cuprates, \cite{Graf} respectively. 
\par
The properties around $(\pi,0)$ are primarily characterized by mode vi$^{\prime}$, which corresponds to the flat band observed in high-$T_{\rm c}$ cuprates and in theoretical calculations. \cite{UniversalFlatbandBi2212,ImadaRMP,DagottoRMP,Kohno2DHub,SakaiImadaPRB,Tohyama2DtJ,Bulut,PreussQP,PreussPG,MoreoQP,Marsiglio1h,MartinezHorsch1h,LiuManousakis1h,Brunner1h2D,Beran1h2D,AvellaCOM,DagottoFlatband} 
Because this mode carries large spectral weights and exhibits an almost flat dispersion relation over a wide momentum range around $(\pi,0)$, 
it significantly contributes to the main peak of the single-particle density of states $\rho(\omega)$ [Figs. \ref{fig:2DtJ}(a)--\ref{fig:2DtJ}(c)]. 
For instance, as shown in Fig. \ref{fig:2DtJ}(p), the energy difference of the main peak of $\rho(\omega)$ from the top of the band at half-filling [$\varepsilon_{\rm p}$] [Fig. \ref{fig:2DtJ}(a)] is primarily determined 
by that of this mode at $(\pi,0)$ [$\varepsilon_{(\pi,0)}$] [Fig. \ref{fig:2DtJ}(a)]. 
%This characteristic can be understood from the flat dispersion relation (saddle points \cite{DagottoFlatband}) and the large spectral weights of this mode. 
%For this characteristic, not only the flat dispersion relation\cite{DagottoFlatband} but also the large spectral weights of this mode will be relevant. 
This energy difference, which can be regarded as a pseudogap in the small-doping limit, will be related to the antiferromagnetic fluctuation 
because it is almost proportional to $J$ in the small-$J/t$ regime [Fig. \ref{fig:2DtJ}(p)], 
as in the 2D Hubbard model. \cite{Kohno2DHub,PreussPG,Bulut} 
\par
Near the Mott transition, because mode vi$^{\prime}$ is located below $\omega=0$, there is no mode crossing $\omega=0$ along $(0,0)$--$(\pi,0)$, and 
the spectral weights along $(\pi,0)$--$(\pi,\pi)$ for $\omega\approx 0$ are also reduced significantly [Figs. \ref{fig:2DtJ}(b), \ref{fig:2DtJ}(c), \ref{fig:2DtJ}(e), \ref{fig:2DtJ}(j), and \ref{fig:2DtJ}(k)]. \cite{Kohno2DHub} 
Thus, the spectral weights for $\omega\approx 0$ essentially remain only around $(\pi/2,\pi/2)$, 
which corresponds to the Fermi arc behavior observed in high-$T_{\rm c}$ cuprates. \cite{ShenRMP,LSCO_FS} 
\par
In the large-doping regime, the behavior becomes similar to that of noninteracting electrons after the pseudogap closes [the peak of $\rho(\omega)$ or the flat mode crosses $\omega=0$] 
[Figs. \ref{fig:2DtJ}(d), \ref{fig:2DtJ}(h), and \ref{fig:2DtJ}(l)]. \cite{Kohno2DHub,PreussPG,PreussQP,Bulut,MoreoQP} 
\subsection{Large negative $\omega$} % ------ Large negative w ------
\label{sec:largeNegative}
Although mode iv$^{\prime}$ corresponds to mode iv, 
its spectral weights around $(0,0)$ and $(\pi,\pi)$ are significantly smaller and larger, respectively, and its $\omega$ values around $(0,0)$ and $(\pi,\pi)$ are significantly higher than those of mode iv 
near the Mott transition [Figs. \ref{fig:2DtJ}(b), \ref{fig:2DtJ}(c), \ref{fig:2DtJ}(e), and \ref{fig:1DtJ}(g)]. 
Similar features have also been obtained primarily at half-filling using exact diagonalization \cite{comHubtJDiag} and recently using CPT independently. \cite{CPTJ3} 
Here, we interpret these features as a consequence of the restriction on spectral weights. 
\par
For the electron operators with the no-double-occupancy constraint, ${\tilde c}^{(\dagger)}_{i,\sigma}$, the spectral weight for each ${\bm k}$ [$n_{\rm t}(\bm k)$] is reduced to $(1+\delta)/2$ [\onlinecite{Stephan_nk}] [Figs. \ref{fig:2DtJ}(f)--\ref{fig:2DtJ}(h) and \ref{fig:1DtJ}(e)]. 
As a result, the spectral weight for $\omega<0$ [$n_-({\bm k})$] around ${\bm k}={\bm 0}$ in the $t$-$J$ model becomes significantly smaller than that of the Hubbard model near the Mott transition [Figs. \ref{fig:2DtJ}(g) and \ref{fig:1DtJ}(h); Figs. \ref{fig:1DtJ}(e) and \ref{fig:1DtJ}(b)]. 
Because the spectral weight for $\omega<0$ ($W_-$) is equal to $(1-\delta)/2$ in the $t$-$J$ and Hubbard models, 
$n_-({\bm k})$ around ${\bm k}={\bm \pi}$ in the $t$-$J$ model becomes larger to compensate for the reduction around ${\bm k}={\bm 0}$. 
%[Figs. \ref{fig:2DtJ}(c), \ref{fig:2DtJ}(i), \ref{fig:1DtJ}(a), \ref{fig:1DtJ}(b), \ref{fig:1DtJ}(d), \ref{fig:1DtJ}(e), \ref{fig:1DtJ}(g), and \ref{fig:1DtJ}(h)]. 
Here, ${\bm 0}$ and ${\bm \pi}$ respectively indicate 0 and $\pi$ for 1D systems and $(0,0)$ and $(\pi,\pi)$ for 2D systems. 
\par
In the RPA from the decoupled-chain limit [$G^{-1}({\bm k},\omega)=G_{\rm 1D}^{-1}(k_x,\omega)-t_{\perp}({\bm k})$], \cite{RPAWen} 
the spectral weights shift upward and downward in the momentum regime for $t_{\perp}({\bm k})>0$ and $t_{\perp}({\bm k})<0$, respectively, \cite{KohnoNP,Kohno2DHub,Kohno2DHubNN,KohnoQ1DH} 
where $G_{\rm 1D}(k_x,\omega)$ and $t_{\perp}({\bm k})(=-2t_{\perp}\cos k_y)$ denote the Green function of a chain and Fourier transform of the interchain hopping integral, respectively. 
In addition, the mode shift becomes large if the mode carries large spectral weights in the large-$|t_{\perp}({\bm k})|$ regime. \cite{KohnoNP,Kohno2DHub,Kohno2DHubNN,KohnoQ1DH} 
Because the spectral weight around $k=0$ for $\omega<0$ in the 1D $t$-$J$ model is significantly smaller, the downward spectral-weight shift around $(0,0)$ is smaller [Fig. \ref{fig:1DtJ}(f)] than that of the Hubbard model [Fig. \ref{fig:1DtJ}(c)]. 
\par
This argument explains why the $\omega$ value around $(0,0)$ of mode iv$^{\prime}$ is higher than that of mode iv near the Mott transition [Figs. \ref{fig:2DtJ}(c) and \ref{fig:1DtJ}(g)]. 
For $(\pi,\pi)$, a similar argument can also explain that the $\omega$ value around $(\pi,\pi)$ of mode iv$^{\prime}$ is higher than that of mode iv. 
Near the Mott transition, however, the $\omega$ values of modes 4, 4$^{\prime}$, iv, and iv$^{\prime}$ around ${\bm k}={\bm \pi}$ appear to be better explained 
by considering that they are almost the same as those around ${\bm k}={\bm 0}$ [Figs. \ref{fig:2DtJ}(a)--\ref{fig:2DtJ}(c), \ref{fig:2DtJ}(e), \ref{fig:1DtJ}(a), \ref{fig:1DtJ}(d), and \ref{fig:1DtJ}(g)]. 
At half-filling, the $\omega$ value of mode 4 at $k=\pi$ is exactly the same as that at $k=0$ because the dispersion relation of the holon mode becomes symmetric with respect to $k=\pi/2$. \cite{EsslerBook,TakahashiBook,Kohno1DHub} 
\par
For the differences in the large negative $\omega$ regime, the neglect of the three-site term is relevant, \cite{comHubtJDiag,CPTJ3} because this term contributes to hopping. 
\section{Discussion and Summary} % ------ Discussion and Summary ------
In this study, the single-particle spectral properties of the 2D $t$-$J$ model for $J/t\approx 0.5$ near the Mott transition were investigated. 
In contrast with conventional exact diagonalization studies, where the spectral weights were calculated only at available ${\bm k}$ and $\delta$ points depending on the cluster size and boundary conditions, 
the spectral-weight distributions for continuous ${\bm k}$ and $\omega$ near the Mott transition with small intervals of $\delta$ were calculated using CPT, 
and how the peaks of the spectral weights form dominant modes and how the modes transform to those of the Mott insulator as $\delta$ gradually decreases 
were clarified in the 2D $t$-$J$ model. 
In addition, through comparisons of the $(4\times 4)$-site CPT results with the $(6\times 6)$-site CPT results obtained by combining the non-Abelian DDMRG method, 
the cluster-size effects were confirmed to be small enough to allow discussion on the overall spectral features. 
\par
Furthermore, the natures of the dominant modes in the 2D $t$-$J$ model were clarified by investigating the $\delta$ and $J/t$ dependences of the characteristic energies and spectral weights 
and by tracing the origins of the modes back to those of the 1D models. 
In terms of the dominant modes, various anomalous spectral features observed in high-$T_{\rm c}$ cuprates, 
such as the doping-induced states, flat band, pseudogap, Fermi arc, spinon-like and holon-like branches, giant kink, and waterfall behavior, \cite{ShenRMP,Graf,kink_Xie,kink_Valla,kink_Pan,kink_Moritz,UniversalFlatbandBi2212,LSCO_FS,XraySWT_PRB,XraySWT_PRL,DagottoRMP,ImadaRMP} 
were collectively explained in the 2D $t$-$J$ model, as in the 2D Hubbard model. \cite{Kohno2DHub}
\par
The results for the natures of the dominant modes 
imply that these spectral features are primarily related to the proximity of the antiferromagnetic Mott insulator, 
which has a low-energy spin-wave mode \cite{AndersonSW} but no low-energy charge excitation, 
and to the existence of different energy scales that characterize the bandwidths of the dominant modes 
rather than to double occupancy, which is completely removed in the 2D $t$-$J$ model. 
\par
Although the spectral features of the 2D $t$-$J$ model in the small-$|\omega|$ regime are similar to those of the 2D Hubbard model, 
significant differences appear in the large negative $\omega$ regime around $(0,0)$ and $(\pi,\pi)$ near the Mott transition 
for the modes which can be considered to primarily originate from the 1D holon modes. 
In this study, the differences were interpreted as a consequence of the restriction on spectral weights. 
\par
Because of the limited resolution, this study could not clarify the properties in the very small-$|\omega|$ regime, 
such as the nature of the excitation in the small-$|\omega|$ limit, the accurate gapless points of the single-particle spectrum, 
and the presence or absence of a long-range order in the ground state, although a superconducting ground state has been suggested in a considerable number of studies for the 2D $t$-$J$ model. \cite{AndersonRVB,DagottoRMP,NagaosaRMP,OgataRev,DagottoSuper,RieraSuper,GrosVMC,YokoyamaVMC,SorellaSuper} 
Further studies are needed to clarify more details and how the anomalous features are related to high-$T_{\rm c}$ superconductivity. 
\begin{acknowledgments}
This work was supported by KAKENHI (Grants No. 23540428 and No. 26400372) and the World Premier International Research Center Initiative (WPI), MEXT, Japan. 
The numerical calculations were partly performed on the supercomputer at the National Institute for Materials Science. 
\end{acknowledgments}
%\bibliography{apssamp}

\end{document}